\documentclass[10pt, oneside]{article}   	
\usepackage{geometry}                		
\usepackage{graphicx}				

\usepackage{amssymb,bm}
\usepackage[export]{adjustbox}
\usepackage{bbold,slashed,hyperref,color}
\usepackage[english]{babel}
\usepackage{hyperref}
\usepackage{float}
\usepackage{enotez}
\let\footnote\endnote
\setenotez{list-name=Endnotes,backref=true,split=section,list-heading = \section*{#1},totoc =section}

\title{First steps towards understanding neutrinos
\\[1ex]\normalsize A tribute to Enrico Fermi on the 90$^{\mbox{\tiny th}}$ anniversary of the $\beta$ decay model
}
\author{Francesco Vissani\\[0.6ex]
\normalsize\em INFN, Laboratori Nazionali del Gran Sasso, L'Aquila, Italy, {\em and}\\[-0.3ex]
\normalsize \em Scuola Superiore Meridionale, Largo San Marcellino, Naples, Italy \\[-0.5ex]}
\date{}							

\begin{document}

\maketitle
\begin{abstract}
We retrace the first steps towards understanding neutrinos, particles predicted by Pauli in 1930 to avoid a supposed violation of time-translation symmetry.  Despite the tendency to reduce the whole story to his intuition and the skill of Reines \& Cowan, according to history 
great strides were made in thirties thanks to precious intellectual tools that combined ideas and mathematics. I refer primarily to the contribution of Fermi, who proposed in 1933 a theory in which matter particles can appear and disappear, prototypical of those at the basis of today's particle physics.  This theory, despite its limitations, led physicists towards the observation of neutrinos, paved the way for further developments - e.g., it anticipated the characteristic of crossing symmetry - and had an impressive scientific legacy. We reconstruct the chain of arguments in the most accessible terms for a modern reader, emphasising the role of theoretical physics and reflecting on some alternative assessments of Fermi's contribution.   
A few technical remarks are collected  in the appendix.
\end{abstract}

\maketitle

\parskip0.9ex

\section{Introduction}
Neutrinos are inextricably linked to the study of symmetries in particle physics.
Their origin is ultimately connected to Pauli's desire not to conflict with the law of conservation of energy~\cite{letteron}.
Throughout the history of particle physics, they have brought us the greatest surprises  of seemingly established principles of symmetry: e.g., 
the symmetry between matter and antimatter, the symmetry of parity, individual lepton number conservation, etc. 
In particular, the fact that neutrino masses are not zero has provided the only experimental evidence available today that the ``standard model'',  
based on the gauge symmetry SU(3)$_{\mbox{\tiny c}}\times$ SU(2)$_{\mbox{\tiny L}}\times$U(1)$_{\mbox{\tiny Y}}$, {\em is not} a complete description of the  world of particles.
Because of these considerations, there is growing interest in their history, and I would like to offer a contribution to the discussion, by taking advantage of the fact that this year marks the 90$^{\mbox{\tiny th}}$ anniversary of the first theory to describe neutrino interactions, due to Fermi \cite{rs33}.

Particle physics  progressed very rapidly in the period in which we are interested, the 1930s~\cite{helmu}, 
and this is particularly true when referring to neutrinos and their interactions.
To express this point of view, one could adopt the same acronym of an important series of workshops dedicated to  ``Weak Interactions and Neutrinos'' (WIN), and talk  of 
the first {\em WIN revolution}\footnote{A second {\em WIN revolution} could be placed in the mid-1950s and concerns the direct observation of neutrinos and the discovery of the chiral structure of weak interactions; on the relevance of the second point, I recommend reading a fine historical essay by Weinberg \cite{wei}. On the other hand, 
some theoretical developments of the 1940s, 
such as the theorisation of new species of neutrinos, or the reasoning on leptonic numbers, 
but even Majorana's breakthrough of 1937 (that we will discuss), 
testify to the continuity of scientific progress in this field.}. 
With these considerations in mind, I would like to  recollect certain ideas formulated in those years, in the most useful and accessible way possible for a modern reader, discussing the limits imposed by the conceptual tools available at the time but emphasising their innovative value.

In the following I will try to accurately assess the significance and effectiveness of the first steps taken in the 1930s, 
comparing them with the road still to be travelled to get as far as science has come today.
As we will be dealing with the understanding of weak interactions and neutrinos, the aspects in the foreground will be mostly 
the theoretical ones.
In this sense the present work differs from others, such as ~\cite{mumup}, which highlights above all the experimental aspects and partly those concerning nuclear physics. A work that offers a broader reconstruction, touching on some theoretical aspects, is that of Pais~\cite{ib}: 
These aspects will be explored in more detail below, highlighting the conceptual and formal foundations of neutrino physics and emphasising in particular the relevance of Fermi's contribution.  

\bigskip
To make the discussion self-contained, I complete this introduction 
presenting in  Sect.~\ref{boh} different opinions on the actual relevance of the contribution of Fermi; 
in  Sect.~\ref{dirac}, the influential ideas of Dirac and other physicists of the time; 
in  Sect.~\ref{na}, the most important model of the nucleus in 1920s before Pauli - the nucleus was the 
main question physicists  were grappling with in thirties. 
This makes it easier to appreciate the way Pauli introduced the neutrino (1930) and to discuss how, over the next three years, his proposal  was modified and reached maturity (Sect.~\ref{nap}). Fermi's $\beta$-decay theory can be considered the crowning achievement of these studies of the nucleus, along with 
the understanding of the neutrino, with its own specific features. 
This is discussed in Sect.~\ref{fermi}, where its influence and fruitfulness is recalled. Finally, we 
return to the historical perspective and critical assessments of Fermi's contribution in Sect.~\ref{millo},
also mentioning certain important advances in the concept {\em neutrino} obtained by Majorana again in thirties.

In order to make the presentation as readable as possible, I will also use some explanatory diagrams, confining the auxiliary 
material in the endnotes and examining the more technical aspects (of interest to a more specialised readership) in Appendix~\ref{apun}.

%

\subsection{Current opinions on the value of Fermi's theory}\label{boh}
When discussing the first steps towards understanding the neutrino and its interactions, Wolfgang Pauli's contribution is always and rightly mentioned; Enrico Fermi's  is also usually mentioned, although sometimes with distinctions and reservations. There is a tendency to emphasise that he gave the particle the name it still bears today - i.e., to recognise the success of a branding - and that he was the author of the first serviceable theory that gives substance to Pauli's ideas. These  attributions are not entirely incorrect but have a limiting character: they suggest that Fermi's contribution  is 
hardly comparable to Pauli's.

Various recent and authoritative works represent similar positions. The list includes~\cite{weiss}, which presents the evolution of field theory without mentioning the first mathematical description of $\beta$ decay, and  \cite{enz}, which discusses the origin of the neutrino concept giving little weight to Fermi's contribution. Similar views are conveyed by the introductory seminars of a valuable conference on the history of the neutrino~\cite{francesi}. A recent example of the same position is in~\cite{machec}, where it is alleged that Yukawa's seminal work was influenced by that of Pauli and Weisskopf (see section 2.4.1 there) but not by that of Fermi (see section 2.3.2 there); a similar impression derives from the book~\cite{bonzo} - see in particular Table 1.2 there.
In the belief that these important works are representative of a convergent attitude to Fermi's $\beta$ decay theory, I would like to return to their stances at the end of this paper (Sect.~\ref{millo2}) after recalling the  relevant facts.

For the while, I simply 
note that different assessments can be found. E.g., 
Gamow devotes an intere chapter of his book
{\em Thirty years that shook physics} precisely to Fermi's theory \cite{30yr};
Max Born, in his text on atomic physics~\cite{mb},  always mentions Pauli and Fermi together when talking about neutrinos or weak interactions; 
the historical accounts of Segr\`e \cite{se}, Amaldi \cite{amuld},
Darrigol \cite{drg} 
or Pais~\cite{ib}   would not seem to offer much support for reductive assessments.
Interestingly, Yang has discussed 
certain reasons  why it is difficult for modern readers 
to fully appreciate Fermi's original work today~\cite{yang}.

 \bigskip
In view of these circumstances, I believe it is not unnecessary to re-examine the 
steps taken by theorists in early thirties, entering into a thorough examination of  ideas, facts and formalism.
It is not just a matter of historical accuracy, but of being able to use them for educational paths and 
assessing their cultural significance. Now, while the published ideas  of Pauli are relatively easy to present, it is not so easy to appreciate Fermi's writings and arguments, even if we succeed to recognise their value: it is a theory based on the best mathematical formalism available at the time, but that differs  
what we are used to today. In other words, it is presented in a different language from the one we commonly employ. In addition, it includes ideas that were revolutionary for the time, yet are in common use today. In order to correctly assess Fermi's contribution in relation to his time, it is therefore necessary to understand the nature of his foundations; in practice, one must take note of the (previous and immediately subsequent) contributions of other physicists, his contemporaries, starting with Dirac, Jordan, Perrin (father and son), Majorana and others.

\subsection{Dirac and his influence}\label{dirac}

Although Dirac participates only marginally in the discussion on the nucleus and neutrinos\footnote{In the current scientific literature, it is very common to use the term `Dirac neutrino'; e.g., from inSpire database, consulted in August 2023, there result 181 articles with this locution in the title, where the first one is from 1980. However, none of these articles refers to any specific work by Dirac on neutrinos, which does not exist, but they all resort to the term in a generic sense: ``neutral particle distinct from its own antiparticle, such as Dirac's electron.'' As we will show below, the first person to consider such a neutrino is indeed Fermi, not Dirac.} (but see below), he had not only a 
profound influence but also a dominant role  in the early 1930s,
in particular, with respect to the new generation of theoretical physicists to which he belonged. This is specially relevant with regard to Fermi; 
see~\cite{kragh,raj-1} for very readable accounts, which includes interesting testimonies from the protagonists of the time. 
Other more personal but perhaps even more interesting accounts, such as from Occhialini's memoirs \cite{memo}, suggest that - while Fermi adhered early to Dirac's views and theories - scientists of the `old guard' such as Rutherford or Bohr, but also Chadwick, were critical about them still in 1932. According to Majorana  \cite{recami,war},  a generalised sense of doubt toward Dirac's anti-electron theory 
persisted in Leipzig at the beginning of 1933\footnote{Majorana's mail  to G.~Gentile Jr.~of June 7, 1933~\cite{recami} records the 
change of attitude in Leipzig.}.
L.~Brown~\cite{brown}  
mentions Landau and Fock in the list of skeptics, and recalls that Pauli  expressed in 1933 reservations on 
the argument used by Dirac for predicting the existence of anti-electrons~\cite{mammamia}.

%

\begin{figure}[t]
\centerline{\includegraphics[width=1.00\textwidth,frame]{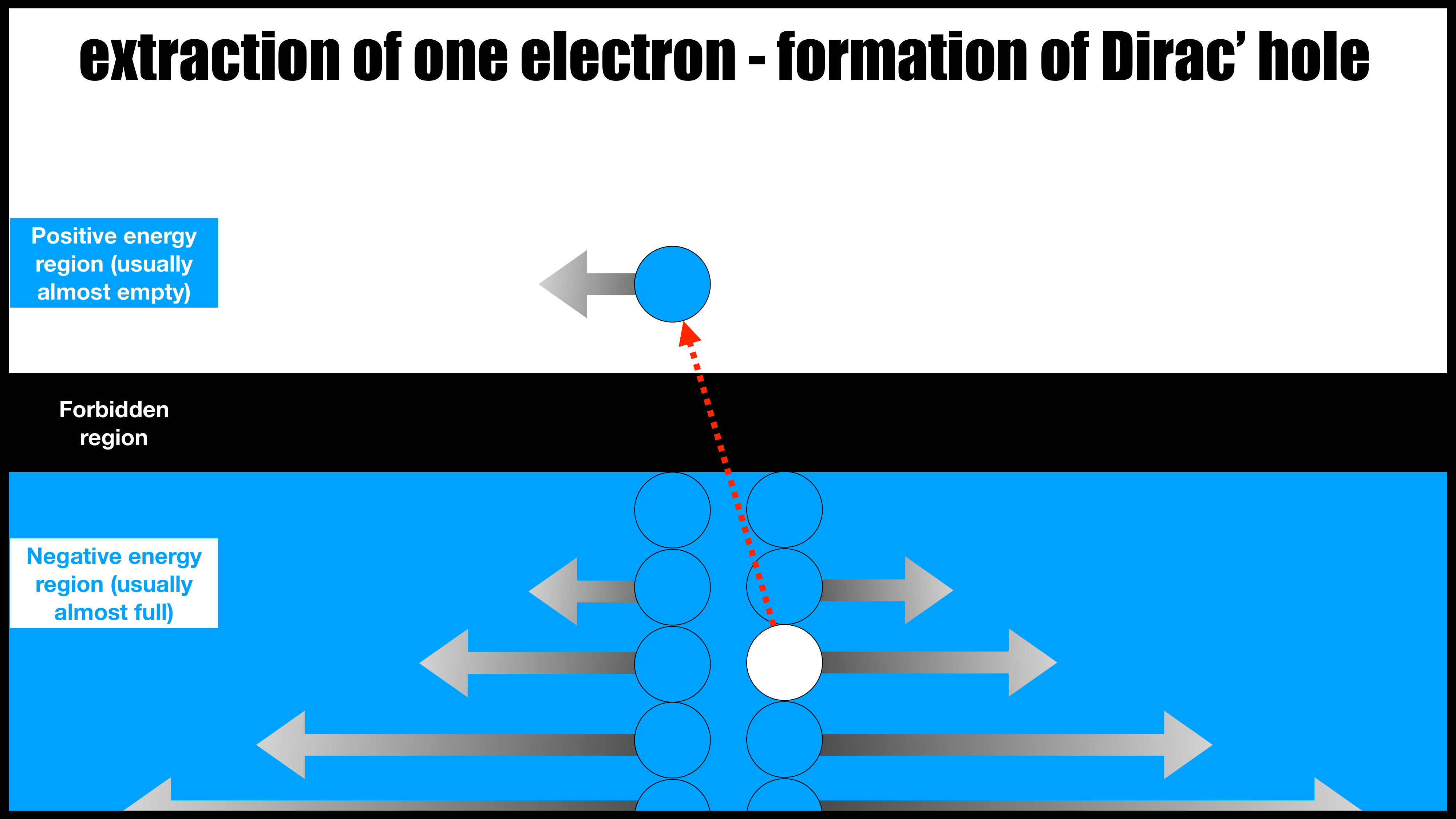}}
\caption{\sf\small Illustration of the key feature of Dirac's 'hole theory'.
In this diagram, the energy states of the electron have an increasing energy from bottom to top and, to highlight this, their possible momenta
$\vec{p}$  are indicated with horizontal arrows for one-dimensional motion; the energies  $-m_e c^2<E<m_e c^2$ are  forbidden,  i.e.,
inaccessible 
 (see Eq.~\ref{spvlg}).
Before extraction from the Dirac sea, all states with negative energy are occupied, gradually descending from the one with $E=-m_e c^2$ 
with $\vec{p}=0$, to those with increasingly negative energies, with longer arrows; after extraction -~caused e.g., by a high-energy photon carrying energy and momentum - an electron  is removed from the sea to the positive energy region.
In short: by supplying a sufficient amount of energy, a hole is formed in the Dirac sea, which appears as a particle with a charge opposite to that of the electron that has been extracted.}\label{fig1}
\end{figure}

\subsubsection{The relativistic wave equation and the Dirac sea}
As is well known, the relativistic wave equation of the 
electron\footnote{See \cite{salvy} for a deep discussion of Dirac's and other wave equations.} (Dirac equation)  
will first lead to a new and more satisfactory 
understanding of spin \cite{de28},  and then \cite{de31} it will 
pave the way for the discovery of the anti-electron (1932) \cite{and}.

This second result came about due to an interesting interpretation of `negative energy' solutions, which corresponds to the 
energies that are admitted by this equation for a freely propagating particle,
\begin{equation}\label{spvlg}
E=\pm \sqrt{ (\vec{p} \, c)^2 + (m_e c^2)^2}
\end{equation}
where  $m_e$ is the mass of the electron and  
$\vec{p}$ its possible momenta.
The interpretation elaborates on a proposal made by Dirac, which aims to avoid inconsistencies between predictions and observations: it consists of the idea that all states of the electron with negative energy exist (consistently with Dirac's equation) but are normally occupied and, in accordance with Pauli's exclusion principle, inaccessible to further particles.
I am of course referring to the so-called `Dirac sea' \cite{de29}, a rather evocative term that I will resort to in the following for maximum 
clarity.  (This concept was perhaps  borrowed from Frenkel's models of electrons in a metal or perhaps suggested 
by other considerations~\cite{kragh}.)
Gamow has made this idea unforgettable with a drawing depicting Dirac, apparently busy thinking... under water and with a fish in front of him~\cite{30yr}.

In Figure~\ref{fig1}, which shows the energy spectrum of electrons, the Dirac sea is depicted at the bottom of the figure. 
All states with negative energies are as a rule occupied; these electrons are assumed to fill the whole physical space. 
Positive-energy states, on the other hand, are all but occasionally free - 
in Figure~\ref{fig1}, there is only one.
In fact, the concept of Dirac sea allows to conceive the idea that, occasionally, some electron of negative energy can be removed from the Dirac sea.
This can happen when sufficient energy is brought by a gamma ray, leading to the formation of a `hole' that is then interpreted as a positively charged particle~\cite{de30},  as schematised in Figure~\ref{fig1}. 
This conceptual scheme, in the terminology of the time, was called `hole theory'.



Figure~\ref{fig1} makes clear that the process just described (nowadays commonly called `$e^+ e^-$ pair creation') was thought of as a simple extraction of an electron from the Dirac sea accompanied by the formation of a `hole', i.e., a process in which the particles always remain the same, before and after, 
or put even more directly, a process in which {\em no creation of matter} takes place\footnote{This is in contrast to today's usage, which is based on later conceptual and formal schemes~\ref{millo1}; e.g., the authors of the current version of Wikipedia~\cite{wikki} speak of ``creation of matter" referring to the 1934 work of Landau and Lifshits~\cite{daus}, where this terminology is {\em not} used.}.
%
%
From this point of view, Dirac and the physicists of his time 
were still contemplating the idea of eternal, unchanging particles; as we will 
emphasize in the last part of section~\ref{na}, the same conceptual framework was the one adopted to describe the 
physics of the nucleus, and to conceive the initial idea of neutrino.

\begin{table}[t]
\begin{center}
\begin{tabular}{l | c| c |l}
Formalism &  Section & Authors & Year \\
\hline
\small Dirac sea based `second quantisation' for QED & \small\ref{dirac} &  {\sf\small  Dirac, Jordan, Klein,}  & {\bf\small  1927-}\\[-0.7ex]  
  & \small  &  {\sf\small  Wigner, Fock}  & {\bf\small  1932}\\[0.7ex] \hline
\small Dirac sea based `second quantisation' for $\beta$ decay  &\small  \ref{fermi}   & {\sf\small  Fermi}  &  {\bf\small  1933} \\[0.7ex] \hline
\small  Dirac sea free `second quantisation'=quantum  &\small \ref{millo1}, \ref{apun3}&   { \sf\small     (Pauli-Weisskopf)} &   {\bf\small  (1934)} \\[-0.3ex]
 \small field theory & \small  &   {\sf\small     Majorana} &  {\bf\small  1937 }\\
\end{tabular}
\caption{\sf\small Three different formalisms to treat relativistic fermions in quantum mechanics.}\label{tabella1}
\end{center}
\end{table}

\subsubsection{Dirac sea based `second quantisation' and its evolution}\label{peppebel}
An equally important contribution by Dirac concerns the formalisation of the way electromagnetic interactions are treated in terms of photons  \cite{d27}; in the long run, this will serve as a seed for the development of modern quantum field theory (QFT).\newline
However, the first fermion quantisation procedure is closely interwoven with the idea of the Dirac sea and will remain so for a long time.
The rest of the 
formalism will be elaborated by Jordan and Klein \cite{jk}; it will acquire a stable form thanks to later work, in particular that of Jordan and Wigner \cite{jw}, and finally of Fock \cite{fo32}, who introduced, already in the title of his article, the naming `second quantisation'. \newline
 The next important step forward, which will show the full potential of the `second quantisation' formalism, 
is the one made by Fermi, that will be discussed in detail. 
Let us recall that Fermi took care to master Dirac's formalism early on, as witnessed by his 1932 review of electromagnetic interactions, that in the second part resumes Dirac equation~\cite{fermirmp}. In his 1933 work on $\beta$ decay, he made original use of the `second quantisation' formalism for fermions based on Dirac sea - or, as he properly called it, the Dirac-Jordan-Klein formalism.\newline 
 These positions will by the superseded in  1937 by the quantisation procedure for fermions 
that is  currently in use (modern QFT) due to Majorana - see Sect.~\ref{millo1} and Appendix~\ref{apun3}. \newline
Table~\ref{tabella1} summarises these evolutions, occurred in the course of a decade or so, stressing the 
major contributors to the issues in which we are interested.

\medskip
In this article, we will use the term `second quantisation' in its original, older meaning; 
we will be interested in the transition between the first two phases, and being interested in relativistic fermions will therefore 
{\em always deal with the idea of the Dirac sea}, 
even when considered just as a technical tool. 
For the reasons just mentioned, one could just as correctly speak of the `old QFT' period, or the `Dirac era' 
to refer to the particular treatment of relativistic fermions used in the early 1930s; 
but we will not need to introduce this new terminology.

\subsection{First model of the atomic nucleus}\label{na}

Finally, we should recall the theoretical ideas on the atomic nucleus, prior to the discovery of the neutron (1932).
We will discuss not one, but two models of the nucleus, which with hindsight we can call `wrong'\footnote{The reasons for this order of exposition are as follows: while the important experimental acquisitions of the early 1900s are usually remembered,  there is a tendency to gloss over the theoretical models that guided the discussion. This type of approach, which would aim to avoid dwelling on `wrong' models, however risks making completely incomprehensible
the expectations or even the results that are then obtained, especially in the cases when these are based on complex intertwining of ideas, formalism and observational results.}: 
the one predominant in the 1920s, described in the rest of this introduction, 
and the one proposed in 1930 by Pauli, which we describe in the beginning of section~\ref{nap}.
In that section, we will then meet a third model of the atomic nucleus, which following the discovery of a new highly penetrating radiation, attributable to particles
which today we call neutron, foresees a nucleus composed of protons and neutrons\footnote{The transition phase is extremely interesting and all but simple; for a thorough review see \cite{ib}, chap.~17, sect.~c. 
It took time to accept the ideas that the neutron is as fundamental a particle as the proton, that the electron is not in the nucleus,  to ascertain the role of $\alpha$ particles, etc,  
see in particular \cite{i2,i} and \cite{solvay}.
We will examine some particularly relevant passages in Section~\ref{nap}.}. Of course, this model (which we owe to Iwanenko, Heisenberg and Majorana) is the basis of the modern one: in its context, the contribution of Fermi to the understanding of the neutrino - Fermi's theory of beta decay - was formulated.


  \begin{table}[t]
  \centerline{
  \begin{tabular}{c|l|c|c}
  model  & year  & ${}^{14}$N  & statistics \\ \hline
  $pe$ & 1920 & $7\times $(2$p$,$e$) & fermion \\
      $pe\nu$ & 1930 & $7\times $(2$p$,$e$,$\nu$) & boson \\
$pn$ & 1932 & $7\times $($p$,$n$) & boson\\
  \end{tabular}}
  \caption{\sf\small Particle content of the ${}^{14}$N nucleus for   different nuclear models and their prediction for the statistics. 
 Its spin was measured and known to be integer, implying  boson statistics\label{tabella2}.}
  \end{table}

\subsubsection{How the first views on the atomic nucleus were forged}

The planetary atom was described already in 1901 by Jean Perrin \cite{p01} and received strong support
 from the famous experiments of Geiger-Marsden, interpreted by Rutherford as evidence of a very small nucleus (1911).
Soon after (actually, already before Bohr's celebrated contributions  to understand the atom \cite{b1}) the internal structure of its nucleus began to be discussed.

The model of the nucleus of that era,
consolidated by various facts and observation, is neatly described by Rutherford (1920) \cite{ruddy}
who operates a synthesis of ideas and results of van den Broek, Moseley, Harkins and Jean Perrin.
Such a model predicted that every neutral atom  comprised as many protons as electrons, 
and some electrons were trapped in the nucleus with the protons.
In this way, besides explaining the neutrality of the atom, and giving support to the emerging Bohr-Rutherford model,
1)~one could explain the fragments of matter that were observed to be 
expelled from the nucleus by some radioactive species, 
namely electrons ($\beta$ rays), helium nuclei ($\alpha$ rays) and later also protons;
2)~the reasoning proceeded exploiting only the known particles;
3)~the gross features of isotopic masses were explained, except for small deviations that could be 
attributed to the details of the nuclear structure.
We will call  it for easy reference 
the $pe$ model, as in the first line of Table~\ref{tabella2}, and we will discuss it further later.

  \bigskip
  Furthermore, it was possible to hazard speculations on bound states between electrons and protons, 
  without any electric charge - which is an anticipation, albeit incomplete and `wrong',
of the modern ideas of the neutron\footnote{See  \cite{ftt} for the very interesting history of the neutron.}; one could think of a nucleus made up of $\alpha$ particles and a few unpaired protons or electrons;
within a few years, it could have been predicted whether the total spin of a nucleus 
was integer or half-integer; etc. 
In order to appreciate the state-of-the-art at the time, we quote a few statements of  Rutherford \cite{ruddy}:
\begin{quote} \sf\small 
\guillemotleft In considering the possible constitution of the elements, it is natural to suppose that they are built up ultimately of hydrogen nuclei and electrons.
[...]
If we are correct in this assumption it seems very likely that one electron can also bind two H nuclei and possibly also one H nucleus.
[...]
Such an atom would have very novel properties. Its external field would be practically zero, except very close to the nucleus, and in con­sequence it should be able to move freely through matter.\guillemotright
\end{quote}
L.~Brown~\cite{lmb} underlines the  doubts and disapproval of Gamow (1931) toward 
the $pe$ model of the nuclear matter \cite{gdis};
another possibly more interesting criticism of a variant of this model appears  {\em after} 
the discovery of the neutron. It is in a 1933 work by Majorana \cite{mui1}
and in translation it reads:
\begin{quote} \sf\small
\guillemotleft
  Heisenberg [...] treats the neutron as a combination of a proton and an electron [...]
One may doubt the validity of this analogy as the theory does not explain the inner structure of the neutron\guillemotright
\end{quote}
see~\cite{pitoz} for more discussion. 
It should be noted, however, that we are talking about the best thinkers of the time; even though it may be hard for a modern reader to believe, it still took {\em several years} to dissolve all reservations about the new point of view\footnote{E.g., 
Enrico Persico, who was informed of the developments, still in  1936 writes in his textbook \cite{purs}
\begin{quote} \sf\footnotesize
\guillemotleft 
It is not yet known then whether these species of particles are all to be regarded as elementary, i.e., indivisible, or whether e.g., the neutron consists of a proton plus an electron, or conversely 
the proton consists of a neutron plus a positron.\guillemotright
\end{quote}}.
%
%

\subsubsection{Comment on the character of the first models of the nucleus}
It should be emphasised strongly and without  hesitation that, with this model of the nucleus, matter particles can combine with each other, 
but are, so to say, eternal.  
In fact, we speak of processes of radioactive `disintegration',  
 i.e.~of separation of parts which were previously integrated among them, namely of a 
 fragmentation: In the model described above, this 
 applies both to disintegration of the $\alpha$ type and to that of the $\beta$ type\footnote{Remember the shock of the chemist Soddy, who realized in 1901 that different chemical elements can transform with each other \cite{soddy}. The observed regularities were then organised by Fajans and him into the `law of radioactive displacements' \cite{fjs1,fjs2}
 (1913). The $pe$ model of the nucleus aimed to proceed beyond the `phenomenological' level of discussion.}.

Rutherford was right, therefore, to recall Prout's hypothesis in this context of discussion; 
indeed, at an objective glance, the adoption of such a model even entailed moving in the direction 
of Greek atomism. Electrons and protons had taken the place of the `atoms' of the Greek thinkers, 
playing, so to speak, the role of microscopic bulwarks of what exists.

The only difference  
 is  in the fact that this model of the nucleus postulated 
two types of fundamental components of matter - protons and electrons -
instead of a single substance, the {\em arche} - along of course with the vacuum. 

In this regard, it is noteworthy that Dirac himself made an attempt to reduce matter to a single substance by trying to conceive of protons as 'holes' in a hypothetical sea of electrons of negative energy \cite{de29,de30}. 
This attempt, later abandoned, was born in the patchwork of ideas and formalism recalled in~\ref{dirac}.

\section{Pauli's proposal and its evolution (1930-1933)}\label{nap}

\subsection{Pauli 1930}\label{cicciob}
Pauli's intervention in the discussion of the $\beta$ decay - i.e, the letter addressed to  the 
`radioactive ladies and gentlemen' \cite{letteron} -
is so famous that it is enough to highlight only the aspects that matter directly for the following.

Severe contradictions had arisen between certain implications of the $pe$ model of the nucleus and the observation \cite{ib,lmb}. 

\begin{figure}[t]
\centerline{\includegraphics[width=1.00\textwidth]{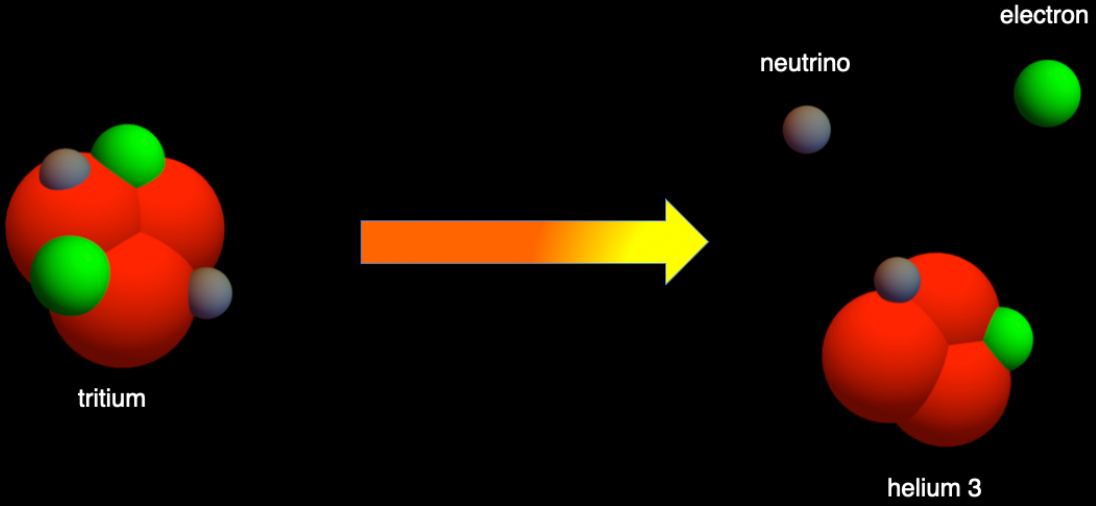}}
\caption{\sf\small The figure illustrates the first idea of the neutrino,  proposed in 1930 by Pauli as part of a theory of the atomic nucleus
and hypothesised to be 
 immutable particles of matter. As discussed in the text, the transition to the modern conception, in which  matter particles  can be created or destroyed just as photons,  took place in the following three years.}
\label{fig2}
\end{figure}

The problems Pauli successfully tackled were the following two: 
\begin{enumerate}
\item[$\star $] the disagreement between the observed and predicted spin of nuclei as nitrogen 14, resumed in  Table~\ref{tabella2};
\item[$\star $]  the fact  that the observed energy of the $\beta$ spectra 
is continuous, not discrete as expected in the $pe$ model.
\end{enumerate}
The devised way out was a modification of the $pe$  model of the nucleus - or better, its upgrade
 into a $pe\nu$ model. 
In fact, Pauli hypothesised the existence of a new particle
inside the nucleus, neutral and almost invisible - see Figure \ref{fig2}. I am referring to the following passage from the above letter,
\begin{quote} \sf\small
\guillemotleft
 I have hit upon a desperate remedy to save the ``exchange theorem" of statistics and the law of
conservation of energy. Namely, the possibility {\em that in the nuclei} there could exist electrically neutral
particles, which I will call neutrons, that have spin 1/2 and obey the exclusion principle\guillemotright 
\end{quote}
(emphasis mine). Thus, the proposal to resolve the two discrepancies between observations and predictions was this: the new particle was assigned a spin of 1/2 and, assuming it was emitted with the electron, was able to carry away a fraction of the emitted energy: See Figure~\ref{fig2} again. 
Note that in the $pe\nu$ model, just as in the $pe$ model, matter particles retain the character discussed in the last paragraph of Sect.~\ref{na}: they remain unchanged before and after the reaction, they are immutable.

%

The statistics of nitrogen could be explained - see again Table~\ref{tabella2} - but three main problems were left unanswered:
\begin{enumerate}
\item[$\bullet$]  the value of the magnetic moment of the nucleus,  
which, as pointed out by Kronig~\cite{kruc}, 
 is always much smaller than a Bohr magneton despite the hypothesised presence of unpaired electrons in certain nuclei;
\item[$\bullet$] the nature of the force that would bind the electrons  to the nucleus; 
\item[$\bullet$] the great kinetic energy of the electrons in the nucleus, attributable to the uncertainty principle.
\end{enumerate}
The typical (often implicit) answer to the first problem was that electrons and neutrinos were bound and not free;  
and later people began to think about a correlation between the spin of the electron and neutrino, perhaps by packing them with the proton within the same particle.
But no neat solution could be found to the last two, apparently related, problems.
However, this was not considered so serious. Indeed,  at the time it was {\em not} taken for granted that 
all principles of quantum mechanics, valid in the atomic regime, were necessarily valid in the nuclear one \cite{ib}. 
This position may be surprising today, but there are countless proofs of this attitude, supported in particular by Bohr: 
In fact, he went so far as to postulate that energy and momentum were not exactly conserved in the $\beta$ decay\footnote{G.~Holton made a very interesting observation in this regard: {\sf \guillemotleft  Bohr is such a bad authority on these [Fermi] papers because Bohr really had it in his mind that there was some profound problem with neutrinos and energy and so on, and didn't want to have it solved except in a mystical and deep way. It was solved by Fermi in `too elementary' a way.\guillemotright} \cite{holton}. This suggests that the origin of the beta decay theory owes something not only to Fermi's ability to do science, 
but also to his personal style.}.
See e.g.,  \cite{30yr,ib} and note incidentally that Gamow enrolls also Ehrenfest in the {\sf \guillemotleft neutrinophobe\guillemotright} party.

Finally, let us remember that Pauli described his ideas publicly but not in a scientific journal;
Although some of his later statements might suggest that this choice was due to personal problems, the impression is that it was inspired above all by {\em reasonable caution.} To ask at the time for an extra particle, and an invisible one at that, seems an extremely bold position. Bohr, as just mentioned, had an alternative explanation for the apparent energy non-conservation in $\beta$ decay; however, there is no need to imagine that Pauli feared a personal confrontation, it may well be that he preferred to stop his proposal at a sufficiently safe point pending further events.

\subsection{Fermi 1932-1933}
Fermi's first documented thoughts on the point appear in July 1932. 
In a report of a popular discussion, he writes \cite{f32,vb,ciokbon}:
\begin{quote} \sf\small
\guillemotleft
One might think, for example, according to a suggestion by Pauli, that there are neutrons in the atomic nucleus that would be emitted simultaneously with the $\beta$ particles. [...]\guillemotright
\end{quote}
The term `neutron' is the one used by Pauli to denote what we now call 
`neutrino', see the previous section.  A note added at the time of printing acknowledges the definitive discovery of the neutron; Chadwick's work had been submitted and published in February 1932. Although one might think that Fermi was clear about the difference, this is not explicitly stated.
On 13 October 1933 (before the 7th Solvay conference) Fermi instead distinguished the two particles and introduced
a new terminology\footnote{This terminology arose from a conversation between the boys in Via Panisperna: see \cite{amuld}. In order to describe a ``light neutron''  the term {\em neutronino}  was  jokingly introduced,  by joining the Italian diminutive suffix -ino. The final term, {\em neutrino}, is nothing but its contraction, and sounds like ``small neutral object''; this word ideally contrasts with {\em neutrone}, which sounds instead like ``big neutral object'', as -one is the Italian augmentative suffix.} suited to describe the new situation \cite{f33,vb,ciokbon}. Here are his words on the subject
%
\begin{quote} \sf\small
\guillemotleft
It would then remain to clarify at a later time the structure of the neutron, for which, as mentioned, the scheme of quantum mechanics probably should not yet be applicable; indeed, we have in this regard, from the $\beta$-ray continuum, some clue which, according to Bohr, would suggest that in these new unknown laws perhaps not even the principle of conservation of energy is valid any longer; when it is not admitted, with Pauli, the existence of the so-called `neutrino'.\guillemotright\end{quote}
At this point, it is clear that   the  words `neutrons' and `neutrinos' mean the particles we know today.
Apparently,  Fermi has become completely familiar with Pauli's ideas, has updated them
to his own time, and started pondering about the structure of the neutron.
His thinking has entered a very interesting transitional phase.


\subsection{The debate  at Solvay 1933}\label{ennio}
At the 7th Solvay Congress in Brussels (held 22-29 October 1933), devoted to
\begin{quote} \em 
Structure and properties of the atomic nuclei 
\end{quote}
Heisenberg gave one of the main speeches, and 
summarized, in the presence of Pauli, Fermi, Dirac, Francis Perrin and others, 
on what was known about the nucleus~\cite{solvay}. 
Here are a few passages from the following discussion, which is extremely interesting.

\subsubsection{Pauli} Pauli \cite{solvay} (as reported also in \cite{enz-libro}) claims as follows, 
%
\begin{quote} \sf\small
\guillemotleft Regarding the properties of these neutral particles, the atomic weights
of the radioactive elements tell us first of all that their mass cannot exceed
much that of the electron.
In order to distinguish them from the heavy neutrons, 
Mr.~Fermi proposed the name ``neutrino''.
It is possible that the
neutrino's own mass might be equal to zero [...]\guillemotright
\end{quote}
His position has become more assertive\footnote{Pauli's position in 1933 was commented by Abraham Pais in a debate at  
\cite{bonzo} (page 275) as follows 
\begin{quote} \sf\footnotesize
\guillemotleft As I understand Pauli, he was a deeply conservative kind of physicist. He wrote a letter, as all of us know, around 1930, about the neutrino. He was asked by Fermi to speak about this at an international meeting, I think at the Marconi meeting in 1931. At that time, Bohr still held the floor about nonconservation of energy, and Pauli, although he was fresh and could really attack Bohr, was also really impressed by Bohr. By the time of the Solvay conference of 1933, Pauli knew on experimental grounds that Bohr was wrong, and then for the first time he opened his mouth.\guillemotright 
\end{quote}
More interesting remarks on the attitude of Pauli are in \cite{lmb}.}. 
Pauli insists that, to satisfy the conservation of energy, the hypothetical particle cannot be too heavy
and therefore it is clear that it is not the particle discovered by Chadwick, which weighs about the same as a proton.
Fermi's distinction is recognised as valid and the new name is approved. Pauli then states that this proposal of his was made in June 1931 in an oral report in Pasadena, of which no written record has  apparently survived.

In a much later work~\cite{p1957} he will add 
that in 1931 he no longer believed that neutrinos were the components of the nucleus;
that he had discussed this with Fermi at a conference in Rome in October of the same year. 
Today, one might be inclined to believe that Pauli disregarded the hypothesis that the neutron is 
a structure composed of a proton, an electron and a neutrino,  but we have no certainty about this from his Solvay's intervention.
Valuable (critical) discussion of the late view of Pauli,  with more testimonies of the thirties,  is in \cite{lmb}.
Note also that, in the  {\em Handbuch der Physik} of 1933 contributed by Pauli 
\cite{mammamia} the hypothesis that the neutron is composed of  a proton and an electron
 is described at pages 797 and 832 by Mott.

%

\subsubsection{Dirac, Fermi, Heisenberg}
In order to  give an idea of where the discussion 
went, we report a few other passages.
At the Solvay meeting \cite{solvay}, Dirac said:  
\begin{quote} {\sf\small \guillemotleft If we consider protons and neutrons as elementary particles, we would thus have three species of elementary particles from which nuclei would be formed. This number may seem large, but from this point of view  two is already a big number.\guillemotright}
\end{quote}
compare with last part of Sect.~\ref{na}
and see  \cite{cipolla} for a discussion  of the various positions  on similar philosophical issues.
Heisenberg, who had meanwhile learned that  Joliot and Curie claimed that some nuclei can emit positrons, observed
\begin{quote} {\sf\small 
\guillemotleft my theory in any case predicts the emission of negative electrons and not positrons.\guillemotright}
\end{quote}
Fermi expresses some doubts, that can be considered the latest echoes of the much more serious problems of the $pe$ model\footnote{The momentum $p\sim \hbar/a$, associated with confinement in a nucleus of dimensions $a$, obeys
$m_e c\ll p\ll M_N c$. Thus, the kinetic energy of the `nuclear electron' of the $pe$ model can be estimated as $E_\beta=p \ c$, whereas 
the kinetic energy of  a nucleon $E_{N}=p^2/(2 M_N)= E_\beta \times E_\beta/(2 M_N c^2)$ is much lower. In other words, the $pn$ model solves to large extent the last problem of  Sect.~\ref{cicciob}.},
\begin{quote} {\sf\small 
\guillemotleft If I prefer to consider (with Heisenberg and Majorana) the nucleus as composed of neutrons and protons
[...] the average kinetic energy of a particle would have to far exceed the energy with which a particle is bound to its neighbours.\guillemotright}
\end{quote}
In response, Heisenberg argued that if most nucleons are arranged in $\alpha$-particles, the problem does not seem so serious\footnote{Although from a logical point of view the problem simply changes level, this answer is reassuring, as the scientific community had by then become accustomed to the existence of $\alpha$-particles. Note {\em en passant} that the way to properly include nucleon clustering effects (such as those implied by Heisenberg's speculation) is still a lively topic in the debate on nuclear models.}.
Many participants mention neutrinos, but this was not the only - and surely not the main - topic of Solvay 7th meeting.

\subsection{Francis Perrin and Iwanenko  on the road of de Broglie}\label{ennios}
The report of Heisenberg mentions \cite{solvay} also a proposal
advanced by Francis Perrin at the Leningrad congress: the $\beta$ 
radioactivity resembles the formation of $e^+e^-$ pairs by the action of a gamma ray,
which suggests the existence of $\beta^+$ radioactivity as well.
This point is reiterated in writing a few months later \cite{cr}, in a work that Fermi mentions, for defending 
the idea that the mass of the neutrino is very small - if not null - with qualitative arguments\footnote{Perrin reasons
of neutrino mass focussing on the {\em maximum} of the $\beta$ decay spectrum. He correctly notes that if the mass of the neutrino and that of the electron were equal, this should lie in the middle of the spectrum - which does not happen - but then proceeds to assume that the neutrino and electron have equal moments in the spectrum maximum - which is not correct. Fermi instead concentrates on the shape of the spectrum  at the {\em endpoint}.}.
A passage 
which I find even more interesting is the following
\begin{quote} \sf\small
\guillemotleft
If the neutrino has zero intrinsic mass, we must also think that it does not preexist in atomic nuclei, and that it is created, as is a photon, 
during the emission.\guillemotright
\end{quote}
Note that similar statements 
concerning the  $\beta$ rays had been advanced as early as in 1930 \cite{ai},  which 
begins  this way: 
\begin{quote} \sf\small \guillemotleft The emission of $\beta$ rays from radioactive nuclei has a certain analogy with the emission of the 
quantum of light from the atoms\guillemotright
\end{quote}
(However, the idea of \cite{ai}
was to extract an electron from the `Dirac sea', which left the unsolved problem of how to model the reabsorption of the `hole' by the nucleus.)
In  the paper \cite{i} by Iwanenko, the first proposing to include  the newly discovered neutron in the nucleus, 
de Broglie is explicitly mentioned.

\subsection{The process of maturation of an idea}\label{ccmcz}
To sum up, we have seen that from 1930 to the end of 1933, a lively investigation of natural philosophy developed, that took advantage of big experimental inputs - remarkably,  the discovery of positrons and neutrons. 
While the ultimate goal was to understand beta decay, the debate touched numerous aspects: the nature of the atomic nucleus, the character of transitions, of particles and of  conservation laws, the existence of new particles, 
etc. In the course of this investigation, the initial idea of Pauli was modified and responded better and better to the demands of theoretical physicists.  
In fact, Pauli started from a conception in which particles of matter - visible or otherwise - were eternal and immutable; therefore they could be occasionally emitted, certainly not created.

Later, some physicists, and noticeably Ambarzumian, 
Iwanenko and Francis Perrin, as remarked by \cite{ib},
began to contemplate a very different scenario. The view they arrived at was difficult to digest as long as one thought of particles of matter as material dots - as pictured in Figure~\ref{fig1} - but became less odd accepting  
 the wave-particle assimilation: in fact, both were 
 guided by de Broglie's vision of a correspondence between the descriptions of photons and electrons~\cite{luigi}.
 Shortly afterwards, the discussion entered a different phase: a full-fledged theory was proposed.

\section{Fermi's theory (1933)}\label{fermi}
\subsection{Conceptual and formal bases of Fermi's theory}\label{cacaficio}

Fermi's contribution was elaborated in 1933; as it is well-known, a version of the work was
submitted to `Nature' but rejected. Nicola Cabibbo reports on unsuccessful attempts to find this writeup, and expresses the  opinion that work of Fermi on the subject had begun {\em at least two months} before the first publication and probably earlier \cite{nicky}. The work was published in the same year in one Italian journal, ``La ricerca scientifica'' \cite{rs33},
with an unassuming title, 
\begin{quote} \em 
Tentativo di teoria dell'emissione di raggi ``beta"
\end{quote}
namely 
\begin{quote} \em 
Attempt at a theory of ``beta"-ray emission.
\end{quote}
 At the beginning of the following year, two 
extensive descriptions of the proposal,  
in Italian \cite{nc34} and German \cite{zs34}, will appear in prestigious journals. 
An accurate English translation will be published only much later, in \cite{w68}\footnote{
The original Italian version \cite{rs33} is cited by the seminal 
papers of Wick and Bethe-Peierls, while Yukawa cites the German version (see Sect.~\ref{cdb}).
We add with a funny bibliometric note: as of June 2023, \cite{rs33,nc34,zs34} and \cite{w68} have received 173, 578, 1590 and 195 citations, respectively.}.

The objective of Fermi's work is to describe the expectations for the phenomenon of emission of beta rays, i.e., of electrons, elaborating on the neutrino hypothesis to reconcile observations and energy conservation.
In modern notations  his proposal can be summarized as follows:
\begin{equation}\label{embes}
(A,Z)\to (A,Z+1)+ e^- + \nu
\end{equation}
or, stated in words: in  the first nucleus a neutron
becomes a proton; 
this is accompanied by the {\em creation} of two particles, that share between them the energy and 
compensate the increased nuclear charge.

This description in a sense takes a step backwards in the discussion: it does not pretend that the model of the nucleus explains {\em why} something happens. This resembles somewhat in spirit Heisenberg's first paper on modern 
quantum mechanics \cite{w1}, which aims to describe observational facts whatever the price. In this case, the price to 
respect the conservation of electric charge and energy is to admit the 
creation of electrons and neutrinos. While this is the same scheme that is used to treat quanta of electromagnetic radiation (photons), it is the first time that matter particles have been hypothesised to suffer a similar fate.

\subsubsection{Treatment of nucleons}
In  $\beta$ decay, a nucleus increases the charge $Z$ by one unit. Fermi accepts  
a)~that the dimensions of the nucleus indicate that the velocity of the nucleons is non-relativistic, 
and therefore one can reason by adopting ordinary quantum mechanics, which was successfully used for the atom;
b)~that inside the nucleus, a nucleon can change state, passing from that of neutron to that of proton, according to Heisenberg's views (isospin) \cite{isom}, which is consistent with  the observed facts and laws of the radioactive decays.
It seems useful to emphasise that transformations between protons and neutrons only involve a change of state of the nucleon, rather than a particle creation or destruction event. 
 In other words, from a conceptual point of view, Heisenberg's ideas fall into the non-relativistic scheme highlighted in the end of Sect.~\ref{na}.
(A modern reader may be surprised that the relativistic formalism is not used directly for nucleons, but Fermi prefers to proceed with caution, in consideration of the fact that the value of the magnetic moments of the nucleons do not match the one successfully predicted by Dirac's theory of the electron\footnote{Let us recall for the sake of clarity that  Pauli's non-relativistic Hamiltonian is compatible with any value of the magnetic moment.}.)


%

\subsubsection{Treatment of light particles}

\begin{figure}[t]
\centerline{\includegraphics[width=1.00\textwidth]{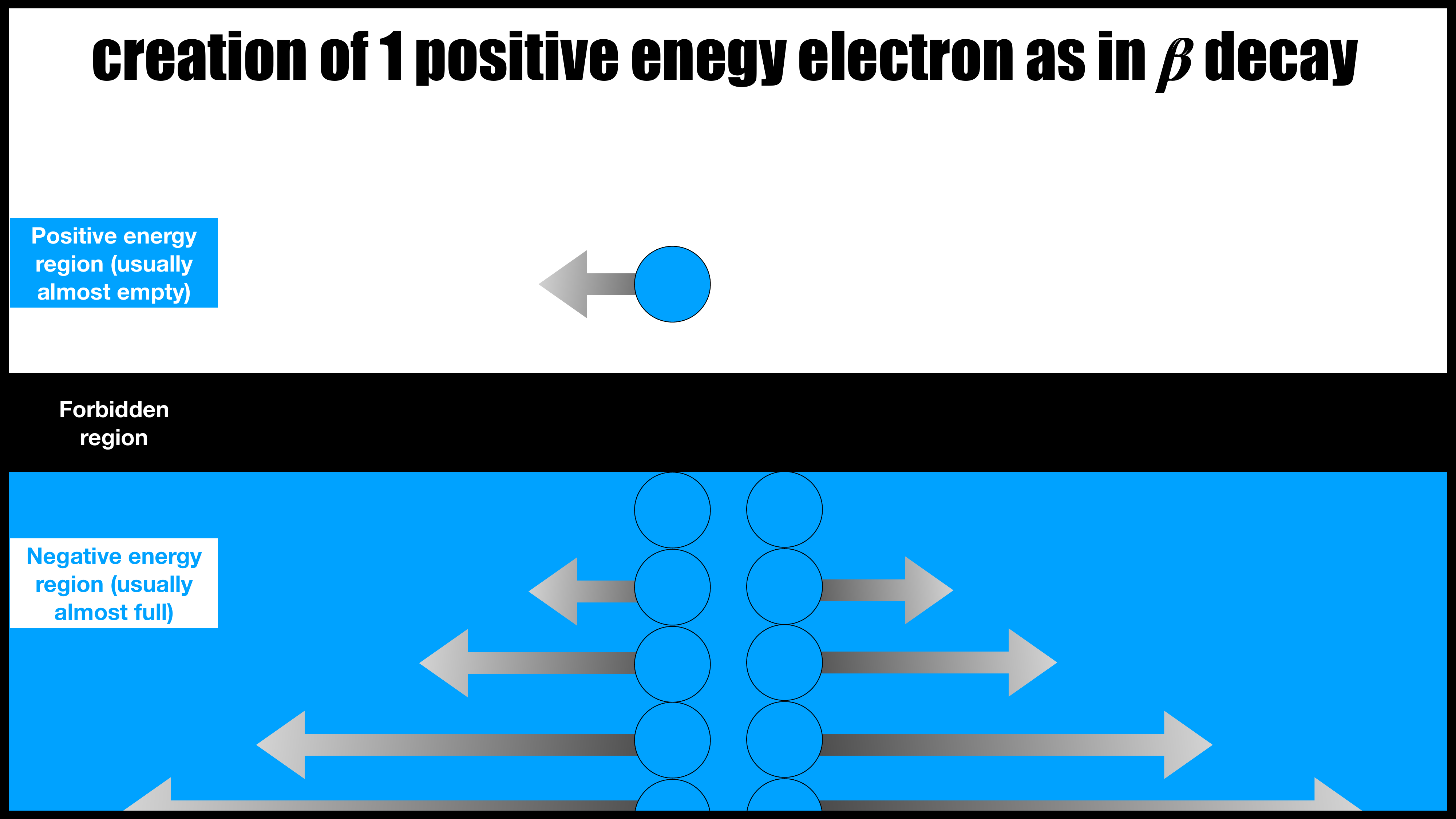}}
\caption{\sf\small The image shows that in the ``second quantisation" formalism, only electrons with positive energy can be created. This allows Fermi to formally describe the processes of electron and neutrino creation, which accompany the increase in electric charge of the nucleus attributed (in accordance with Heisenberg's isospin ideas) to the change of state of a single nucleon - from neutron to proton.
The arrows schematically indicate the possible states of motion of the electron, exactly as in Figure~\ref{fig1}.}
\label{fig3}
\end{figure}

For the purpose of the present discussion, the most interesting aspect 
is the way in which the light particles are treated. First of all, Fermi notes that the relativistic Dirac-Jordan-Klein formalism, what we have called `second quantisation' above, allows us to contemplate the possibility that matter particles of positive energy are created. This is a very innovative position, used to describe the emission of an electron and a neutrino in the final state of a beta decay process\footnote{One might ask why such reluctance to accept this, when the possibility of light quanta being absorbed/emitted had been considered since 1905~\cite{eisto}. But we should not forget that the full acceptance of the photon idea took about 20 years. Apparently,   it was hoped that matter particles (unlike radiation) were immutable, cf.~Sect.~\ref{ccmcz}.}. 
Considering electrons as an example, the proposed description is that illustrated in Figure~\ref{fig3}.
The quantum field of the electron ${\bm \Psi}$, according to this formalism, is simply
\begin{equation}{\bm \Psi}= \psi_s\ \bm{a}_s
\label{capperi}
\end{equation}
The Einstein convention of repeated indices is adopted to sum  over 
 all the possible states $s$, with positive {\em and} negative energies,  
 the corresponding  wavefunctions  $\psi_s$ and   the 
step operators $\bm{a}_s$. This operator has non-zero matrix elements 
only between an occupied state of electron and the vacuum; stated otherwise, 
it describes the possibility that a pre-existing electron is annihilated (disappears) by 
the effect of an interaction. Therefore, its hermitian conjugate conversely describes the possibility that an electron is created.  (See Appendix~\ref{apun1} for details.)

Dirac sea is invoked to exclude that particles of negative energy can be created.
Fermi's make this important point explicit in the publication of 1934; here is 
his explanation in translation~\cite{w68}
\begin{quote}
{\sf\small \guillemotleft Only positive eigenvalues are to be considered. Negative eigenvalues are eliminated by an artifice analogous to Dirac's hole theory.\guillemotright}
\end{quote}

It is important to emphasise that
\begin{enumerate}
\item these ideas give substance to the views on $\beta$ decay,  emerged from the discussion of 
Pauli, Fermi himself, Ambarzumian, 
Iwanenko and Francis Perrin and discussed Section~\ref{nap};
\item  Fermi strongly emphasises that his theoretical scheme differs from that of electromagnetic interactions.
Using the language of the two previous sections, the formation of $e^+e^-$ {\em should not} be seen as a process of creating particles of matter, but simply as the extraction of an electron from the Dirac sea (the formation of a hole); on the contrary, the process of Eq.~\ref{embes} 
describes  the creation of 2 particles of matter\footnote{A common framework for interpreting these two processes will be later recovered, first thanks to an appropriate definition of the leptonic number, and then  through a full development of the idea of gauge bosons of weak interactions (continuing on the path that will be started by Yukawa, see next section).};
\item  for the theory of $\beta$ rays, the role of the Dirac sea is 
simply to guarantee the stability of the world, as regarded in Dirac' theoretical scheme   (Figure~\ref{fig3}), 
 rather than the role of container for particles, as in the hole theory   (Figure~\ref{fig1});
\item  Fermi  refers to Dirac sea with the word {\sf \guillemotleft artificio\guillemotright}, a word which in its English version, `artifice', suggests the sense of deception but is more ambiguous in Italian, since the original sense (mediated by Latin)  recalls the word `artistry' and  
indicates an ingenious mathematical technique\footnote{Kragh claims that {\sf \guillemotleft Fermi was~[...]~skeptical of Dirac's theory\guillemotright} (see~\cite{kragh} pag.~114) 
but this  statement is not elaborated. To be sure, the words about the `hole theory' quoted in the text just after Eq.~\ref{capperi} 
(from the English text~\cite{w68}) do not denote excessive consideration towards Dirac sea, 
but the fact remains that Dirac sea is a {\em cornerstone} of Fermi's theory. Compare also with Sect.~\ref{millo}.};
\item since the formalism of `second quantisation' is also used for neutrinos, Fermi assumes the existence of a different Dirac sea for these particles. Therefore, despite the fact that they have no electric charge, neutrinos are somehow distinct from their own antiparticles in his scheme; this type of neutrino could legitimately be called `Fermi's neutrino', but in modern terminology it is usually called `Dirac's neutrino'.
\end{enumerate}

\subsubsection{More facets}
Fermi - who was aware of the wide latitude of possibilities - choses and proposes a specific Hamiltonian, discussed in 
Appendix~\ref{apun2}. 
On this basis, he predicts the energy distribution of the electrons produced: this will be adapted over time - rather than radically changed - and will usefully guide the debate in the years to come.
Quite interestingly, Guerra and Robotti  \cite{gr0} remark that {\sf\small \guillemotleft
These successes take place along lines of research that completely subvert the schedule provided above [...] Some merit to this success should surely be attributed to Majorana\guillemotright};  the first statement refers to initial research plans of Italian physicists, the latter to Majorana's results on neutrons mentioned above~\cite{mui1} and  advertised by Heisenberg~\cite{solvay}\footnote{It could also allude to the speculation that Majorana favoured the dissemination in Rome of the thought of another German scientist, Jordan. Other (non-exclusive) possibilities are that Fermi learned about Jordan-Klein from the literature, or from some of his other collaborators, or from Heisenberg himself in Brussels. If a pun is allowed, I think  one should admit that Fermi was quite interested in the description of fermions.}.
The reader who wishes to follow the formal elaborations   can consult the appendices~\ref{apun1}
and \ref{apun2}, and this should make it easier to read it in the original.

In his paper, Fermi also discusses a)~the smallness of the neutrino mass, 
b)~the effects of the wave functions of nuclei,  
c)~the distortion of the electronic plane waves (caused 
by the proximity of the nucleus), 
d)~the possibility of forbidden transitions,   
and other aspects as well. Since the  main purpose of these notes is to appreciate the ideas and formalism underlying 
of Fermi's theory, we will not go into these important aspects.
Instead we  comment  in the next section on the direct influence of this work,  if only to take in how valuable Fermi's paper has been in anticipating and preparing current theoretical conceptions.

\subsection{Impact   on the research at the time}\label{cdb}
To illustrate the importance of Fermi's article, let us recall some well-known theoretical advances which appeared soon after and which respectfully cite his work. We focus on three papers presented in 1934, their authors being Wick (4 March); Bethe and Peierls (7 April); Yukawa (17 November).
For more discussion, see~\cite{amuld}.

\subsubsection{Other decays and electronic captures}
Wick, who was Fermi's assistant since 1932,
derived two further predictions \cite{wick}: first, the reverse process
$(A,Z+1)\to (A,Z)+ e^++\bar\nu$ in which a positron is emitted ($\beta^+$ decay), provided it is compatible with energy conservation,
which explained certain observations of Joliot and Curie\footnote{Other interesting theoretical advances, subsequent to Joliot and Curie's discovery but not directly related to the formulation of Fermi's theory, are recounted by Francis Perrin in~\cite{fp2}. Compare also with Sections~\ref{ennio} and especially~\ref{ennios}.}.
This process is thought of as the conversion of a proton into a neutron inside a nucleus, accompanied by the destruction of an electron and a neutrino with negative energy, i.e., by the formation of two `holes'.
Furthermore, the model of Fermi made it possible to imagine the case in which, instead of forming an electron hole, an initially present electron is `destroyed', according to the scheme: $e+(A,Z+1)\to (A,Z) +\bar\nu$. 
This process, now called `electron capture', will be distinctly observed a few years later by Allen \cite{allen} (1942).
(For some formal detail, see Appendix~\ref{apun3}.)

\subsubsection{Neutrino interactions}
Bethe and Peierls \cite{bp}, referring to Fermi's work in 1933, emphasised the importance of the new conceptual framework involving an electron and a neutrino. They also observed that this leads to the prediction of $\beta^+$ decay and showed that neutrinos {\em must interact.} Their estimate of the cross section
 $\nu_e+ p\to n +e^+$, was simple and witty, being based essentially on dimensional considerations,
 and in essence correct. 
 Its minuscule value, however, led them to claim that neutrinos were undetectable, a pessimistic view that had lasting effects, although it was disproved within about 20 years \cite{rc}. (Interestingly, this work was published in {\em Nature,} 
as well as another  paper of comment \cite{mcr} which was somewhat critical and 
in hindsight had less merit. However, in 1938, the same journal 
had  saluted the value of Fermi's work in written form~\cite{rtv}.)

\subsubsection{Investigation of the nature of interactions}
Yukawa  
  was interested in better understanding the nature of interactions between particles, and in particular what happens   among the particles contained in the nucleus\footnote{It is interesting to compare his motivations and those of Igor Tamm and Iwanenko \cite{bildung}.}. 
  The first two citations in his paper \cite{yuk} are for Heisenberg and Fermi:
 the  former considers important  the `exchange' interaction between protons and neutrons, and finds support for this line of thought in the theory proposed by the latter. Aiming to mimic the description of electromagnetic interactions, Yukawa postulated that an energy quantum is emitted in the exchange interaction (a well-known circumstance) and assumed that the coupling between nucleons $ g $ is much greater than that between light particles $g' $. He developed his theory of $\beta$ decay, which in his intentions modifies and completes that of Enrico Fermi; the adopted notations are almost identical.

\medskip
With hindsight, we can attribute the first two results to ``crossing symmetry'', which is usually considered a typical feature of quantised field theory.  
But this theory was  still to come, and  even the terminology ``cross (or crossed) diagram'',   will be introduced
20 years later \cite{gimmy}.
It should then be recognised that these papers written in 1934  
anticipate this important feature,
despite the `primitive' context of the discussion, i.e., what was called `second quantisation' at the time.
Concerning the latter paper, the value of the  ``Fermi constant'' calculated by Yukawa turns out to be proportional to $g$ and $g'$ and inversely proportional to the mass of the Yukawa quantum squared, an expression that is very reminiscent of the one obtained in the current ``standard model''
of particle physics (electroweak theory).

To summarize, and without taking anything away from the merit of these authors, it is certainly possible to admit that Fermi's work had the effect of stimulating valuable scientific developments.

\section{Further developments and discussion}\label{millo}
In this section we address two last important topics. Firstly, we complete the historical picture by examining some formal developments from the 1930s that would eventually lead to replace the `second quantisation' with `quantum field theory', rendering the use of Dirac's sea concept obsolete, moreover introducing a new and deeper idea of  the neutrino (Sect.~\ref{millo1}). Then, since we have completed our overview of the main historical processes that shaped the concept {\em neutrino}, we will ready to return to the alternative evaluations of Fermi's theory presented in the introduction (Sect.~\ref{millo2}).

\subsection{On the way to modern formalism}\label{millo1}
The Dirac sea based `second quantisation' procedure 
was felt as unsatisfactory by certain theorists,
such as Bohr or Pauli \cite{kragh}
but it secured anyway  certain significant scientific successes (Dirac, Fermi, etc). 
It has two radically asymmetric features: the Dirac sea, and 
the form of the fermionic field (eq.~\ref{capperi}).
These compensate for each other yielding a laborious but ultimately  usable 
procedure. Indeed as 
we have just seen, this formalism features crossing symmetry, 
which we know to be a major characteristic of the subsequent quantum field theory as well.

Overcoming the Dirac sea concept will take about 10 years. 
A first step is the one performed by
Pauli and Weisskopf for scalar particles in 1934 \cite{pw}, to which Pauli
- according to the testimony of the second author \cite{weiss} - was fond of referring to as the `anti-Dirac theory'.

For fermionic particles (those of Fermi's  theory of $\beta$ decay) the decisive step forward will be done by Majorana, who wrote a famous work eloquently entitled
{\em A symmetric theory of electrons and positrons} \cite{maj}.
From 1937 on, all the modern elements of the theory of quantised fields exist\footnote{See again~\cite{gr0}  for a more useful discussion. Esposito pointed out an interesting previous result in Majorana's textbook~\cite{wars} on the possibility of quantizing the Klein-Gordon (scalar) field. A memoir by GC Wick in the Amaldi Archive, also recalled in~\cite{wars}, claims that Majorana was aware of this from the time of the Rome conference in 1931.}.
Fermi, who was commissioner for the competition  in the same year \cite{recami,salvo}, 
wrote these words of comments in the  judgment for Majorana:
\begin{quote} {\sf\small\guillemotleft
he devised a brilliant method that permits us to treat the positive and negative electron in a symmetrical way, finally eliminating the necessity to rely on the extremely artificial and unsatisfactory hypothesis of an infinitely large electrical charge diffused in space, a question that had been tackled in vain by many other scholars.\guillemotright

} 
\end{quote}
Comparing with Sect.~\ref{cacaficio}, we see that Fermi's position towards the Dirac sea has become much 
more critical than in 1933.
A last formal generalization will be made
by Kramers \cite{kramers}, who
in his work, remarks:
\begin{quote} \sf\small\guillemotleft
we will in this article represent some results of the Majorana~calculus (and, thereby, also of the Dirac-Heisenberg formulation of the hole theory, with {\em which it is practically equivalent}) in a general form\guillemotright
\end{quote}
(emphasis mine.)  
This statement suggests that abandoning Dirac's sea to quantise spin 1/2 particles is primarily a matter of convenience, which is correct for the electron\footnote{Dirac went so far in his speculation as to deny the possible existence of matter particles with spin other than 1/2, based on a supposed ``necessity" to postulate the Dirac sea to avoid states with ``negative energy." Pauli and Weisskopf's construction reveals that this is not true: particles with spin 0 can be consistently quantized. Majorana's has an even more radical effect on Dirac's position, showing that it is based not on facts but on a hypostatization of a mental construction: there is no need to talk about negative-energy states.
Only at this point in the discussion (7 years after \cite{de29}) were all the theorists able to recognize the weakness of Dirac's position.} and a large number of cases of interest-all charged particles-but there is an important exception, which Kramers misses. In fact, hole theory requires that particles and antiparticles be different, a circumstance that is not met in the case of Majorana's neutrinos. It is worth mentioning that this kind of neutrino is not considered  an academic construct, but a rather plausible hypothesis that engages many current experiments in its search.
For details and discussion see Appendix~\ref{apun3}.
%
%


\subsection{On the historical significance of Fermi's theory}\label{millo2}
%

As we saw at the beginning of this discussion, the literature records several reservations about the actual relevance of Fermi's contribution. In this final section, we propose to return to the cases cited at the beginning, which seem to us to be representative of such positions.

Perhaps the most interesting case is that of Weisskopf  \cite{weiss}; but since Fermi is not mentioned at all, all we can do to discuss the significance of this omission is to advance a few conjectures.
It can be assumed that Weisskopf deems relevant that Fermi treats fermions using the `second quantisation' (Dirac-Jordan-Klein's) formalism  
and not the modern quantum field theory. In support of this conjecture, we cite the extensive critical discussion of the Dirac sea in~\cite{weiss} and the fact that much emphasis is placed on the work~\cite{pw}.  
Weisskopf formal point of view most likely reflects that of Pauli\footnote{However, it must be made clear that making the assumption of a reservation of a formal nature does not mean that the value of Fermi's paper was not appreciated; this is consistent with an account reported by Rhodes \cite{rhodes}, who quotes Weisskopf's own comment: {\sf \guillemotleft  A fantastic paper [. . .] a monument to Fermi's intuition.\guillemotright}
See also \cite{wwp}.}. 
But while today it is natural for us to consider the use of quantised field theory as desirable feature,   it is not obvious that this  is enough to deny the enduring phenomenological value of Fermi's theory (on which it is unnecessary to insist) nor, from a theoretical point of view, the fact that he devised the 
first formal description of a process where  new particles of matter are created. 
A passage of a 
conversation between Wigner and Yang (reported in~\cite{yang}) is quite illuminating in this respect
\begin{quote} \sf\small
{\em Wigner:} 
\guillemotleft Von Neumann and I had been thinking about $\beta$-decay for a long time, as did everybody else. We simply did not know how to create an electron in a nucleus.\guillemotright\\
{\em Yang:} \guillemotleft Fermi knew how to do that by using a second quantised $\psi$?\guillemotright\\ 
{\em Wigner:} \guillemotleft Yes.\guillemotright\\
{\em Yang:} \guillemotleft But it was you and Jordan who had first invented the second quantised $\psi$.\guillemotright\\
{\em Wigner:} \guillemotleft Yes, yes. But we never dreamed that it could be used in real physics.\guillemotright
\end{quote}



Enz, in~\cite{enz}, expresses surprise that Pauli did not himself formulate the theory that we owe to Fermi. But as we have just mentioned, Pauli had not even considered the idea of using the `second quantisation' procedure; 
he was probably more focused on devising a formal alternative to this scheme. (Instead Fermi had already moved in that direction in his works on electromagnetism \cite{fermirmp} as recalled above.) 
In the valuable collection of writings of Pauli, edited by Enz and von Meyenn~\cite{enz-libro}, 
we find another precious testimony of Viktor Jakov Frenkel
\begin{quote} \sf\small
\guillemotleft Pauli was critical of Fermi's theory of $\beta$ decay and discussed how inappropriate it was to apply perturbation theory to describe this phenomenon and, accordingly, to develop the quantum mechanical equation according to powers of a small parameter (the Fermi constant) \cite{pauli38}\guillemotright
\end{quote}
While the remark testifies to Pauli's critical attitude, it is now clear to every student that Fermi's is the prototype of an {\em effective} theory, which is to be used at the first order in perturbation theory and furthermore (by definition) is not to be used for higher orders. Therefore, the criticism misses the point, and it is a fact that Frenkel himself (who adds many other interesting points) admits that Pauli did not follow up on this criticism~\cite{enz-libro}.
Finally, even if Pauli himself declared in 1957 that he had soon abandoned the idea of neutrinos in the nucleus, we do not know how far he had arrived in 1933 in the elaboration of this idea.


On the choices in~\cite{francesi}, we note that Allan Franklin's contribution is entitled
{\em The prehistory of the Neutrino.} It concludes with Pauli's letter, which, as we have discussed at length, describes a different concept of neutrino from the current one, closely linked to the old models of the atomic nucleus and to 
ontological conceptions of particles that Fermi's theory would later lead to abandon. The next contribution,
edited by Cecilia Jarlskog, covers the period up to the discovery of Reines and Cowan. It is stated that Fermi's theory 
{\sf\small \guillemotleft
is doomed to fail at high energies\guillemotright}, conceding that this does not make it less useful at low energies
(which agrees with the previous assessment of Frenkel), 
and acknowledging that Fermi's work starts a {\sf\small \guillemotleft global neutrino fever\guillemotright}; then, 
the importance of Francis Perrin's observation (prior to Fermi's work) is strongly underlined.
We have no particular observations to add, except to express the opinion that the realisation of a mathematical theory (=of a formal system), such as Fermi's, has no automatic character;  at the time, this was immediately recognized e.g., by Iwanenko, who spoke of `Fermi formalism' \cite{bildung}. Furthermore,
as witnessed by history, this step has allowed great progress, and has allowed us to predict a series of processes involving the neutrino that had not previously been considered and which were then confirmed. In other words, and resorting to a paraphrase, it seems to us that Fermi has some merit - not only success -
if from the prehistory of the neutrino we have passed to its history. Incidentally, this 
coincides with a statement one finds in the review~\cite{riv}, that after recalling the publication of \cite{rs33}, remarks:
{\sf\small \guillemotleft Here began the history of ``weak interactions.''\guillemotright}


Next, let us examine \cite{machec} from which we cite a passage:
\begin{quote} \sf\small
\guillemotleft the quantisation of the charged Klein-Fock-Gordon field by Pauli and Victor Weiskopf {\em [sic]} in 1934 demonstrated the possibility of pair production without the Dirac sea. It was in this theoretical framework that Yukawa developed his meson theory\guillemotright
\end{quote}
It is not impossible that the alleged influence occurred: Pauli and Weisskopf's work \cite{pw} had appeared two months before Yukawa's meson theory was submitted for publication, and today it would be natural to think that scalars cannot be quantised in the manner of Dirac. On the other hand, it is a fact that Fermi's work is cited directly by Yukawa (in the manner already mentioned), while that of Pauli and Weisskopf is not. Indeed, this consideration entitles us to think that Fermi's adoption of the `second quantization' procedure was not an insurmountable obstacle for Yukawa to appreciate its value, nor that it automatically deprived him of a valuable source of inspiration.

Finally, we consider the influential book \cite{bonzo},  
based on the lectures and round tables of the International Symposium on the History of Particle Physics 
held at Fermilab in May 1980.
In  Table 1.2 there, Yukawa is mentioned first, then Pauli and Weisskopf, and only in third place Fermi; a careful reader should ask why the {\em reverse historical order} was chosen in a history book. 
The Editors, in their introduction, make it clear that 
\begin{quote} \sf\small
\guillemotleft
The theoretical underpinnings were addressed by Paul Dirac, Victor Weisskopf and Satio Hayakawa\guillemotright
\end{quote}
and it is a fact that Fermi's theoretical contribution does not receive much prominence in these three reports\footnote{This is even more true in the case of Majorana, whose contribution is completely ignored.}.
Fermi's work is repeatedly cited by several authors of this book, but  in  
the subject index, there are no locutions such as `Fermi interaction', `Fermi's theory of beta decay' and `Fermi constant'.
Summarizing, this book is valuable and useful in several respects, but it seems to us that its main objective was to collect the oral testimonies of some of the last protagonists of the past, which does not necessarily coincide with that of reporting the written testimonies of the time, nor with that to attempt their  comparative analysis.


\section{Summary}
In 1930, scientists knew of only two particles of matter: electrons and protons. To propose that a third existed, only to explain phenomena relating to the mysterious nucleus of the atom, even with all the caution of the ways in which it was done by Pauli, required enormous intellectual courage. In the years that immediately followed, the neutron and positron were discovered, and the ideas of the nucleus, but also of the $\beta$ ray and of the neutrino, were modified and matured. 

It was Fermi who put all the pieces together, in a model of $\beta$ decay that involved all matter particles known at the time. The intuition of Iwanenko and Francis Perrin, that the emission of matter particles could be likened to that of photons, was linked by Fermi to Dirac's relativistic theory of the electron, and the formalism of Jordan et al.~was used for the first time to describe the creation of matter particles. 

This formalism, the one used to present Fermi's theory, has now been abandoned and there is no doubt that the $\beta$ decay model has evolved over time in various aspects; but these facts have not in the least affected the innovative potential of Fermi's theory. Enz, in a beautiful comparison \cite{enz},
recognises that Pauli's attitude towards the principle of energy conservation 
was 
\begin{quote} \sf\small\guillemotleft
an almost mystical belief in the harmony of the world, similar to Kepler's\guillemotright
\end{quote}
After 90 years, and in the light of history,  Fermi's skill and  determination in constructing a working theory of $\beta$ decay, organising into an organic vision all the best  theoretical tools available at the time, 
pushed to their limits, 
 gives the impression of having been  informed by  a similar passionate and  far-sighted  vision.
 
Moreover,  as  we have repeatedly emphasized, 
Fermi's theory led us to definitively abandon the idea of permanent (eternal) particles of matter. 
Subsequent attempts to recover the reassuring point of view from which we started - which dated back to Greek atomism (Sect.~\ref{na}) - by means of empirical and abstract laws, such as that of the conservation of leptonic number, will almost inevitably lead us to question what exactly is the degree of their conservation, thus bringing us closer to current conceptions of particles of matter (see Sect.~\ref{millo1} and \ref{apun3}). 

We close with a comment of an educational nature, 
inspired by comparing the simple and transparent expression for quantised fields Fermi used, Eq.~\ref{capperi},
with the more complete but also much more complex expressions in use today. One finds oneself thinking that a good teaching effect could be achieved in university lectures if, instead of immediately presenting the finished product of many years of reflection, one spent some time recalling certain passages in the history of fermionic field quantisation.

\subsubsection*{Acknowledgments}

{\small
Work prepared with partial support from research grant 2022E2J4RK {\em PANTHEON: Perspectives in 
 Astroparticle and Neutrino THEory with Old and New messengers} 
 under the PRIN 2022 programme funded by the the Italian 
``Ministero dell'Universit\`a e della Ricerca'' (MUR).

Presented at the  Symposium `Passion for Physics 2023'  of the Italian Physical Society (SIF)
on the occasion of the   70$^{\mbox{\tiny th}}$ Anniversary of the International School of Physics ``Enrico Fermi",
Varenna (Lake Como), June 2023,   and   at the XLIII National Congress of the Italian  Society   for the History of Physics and Astronomy (SISFA), Padua, September 2023.

I thank  
Sanjib Agarwalla,
Giorgio Arcadi, 
Angela Bracco, 
Olivier Darrigol,
Antonia Di Crescenzo, 
Ilia Drachnev, 
Salvatore Esposito,
Giuliana Galati, 
Giovanni Grilli di Cortona, 
Luciano Maiani, 
Nello Paver, 
Esteban Roulet, 
Marco Segala and 
Fred Wilson for stimulating discussions, 
and the reviewers for valuable comments. 

}

\newpage

\appendix

\section{Formalism for treating fermions}\label{apun}
\subsection{Relativistic particles in Fermi's `Tentativo'}\label{apun1}

\begin{table}[t]
\begin{center}
\begin{tabular}{c | c | c| c | c | c }
      &   second  &     label &   quantized &   wave-  &  specific\\[-0.4ex]
       &   quantized field &    of state &   amplitude &   function  &  Dirac sea \\[0ex]
\hline
electrons & $\bm{\Psi}$ & $s$ & $\bm{a}_s$  & $\psi_s$   & $ | \mbox{sea}_s\rangle $ \\[0ex] \hline
neutrinos & $\bm{\Phi}$ & $\sigma $ & $\bm{a}_\sigma$  & $\phi_\sigma$ & $ | \mbox{sea}_\sigma\rangle $  \\[0ex] \hline
\end{tabular}
\caption{\sf\small Notations to describe relativistic fermions in Fermi's theory.}\label{tabella3}
\end{center}
\end{table}

\subsubsection{An original usage of the `second quantisation'}
The starting point is a `second quantisation' formalism\footnote{We reiterate for clarity that throughout the text, we denote (following Jordan) by `second quantisation' the procedure that applies to particles with spin 1/2, relying on the idea of Dirac sea.  This is the procedure used by Fermi.  
We distinguish it from `quantum field theory', the modern formalism, 
which instead is of general applicability -  -  see Sect.~\ref{peppebel} (in particular Table~\ref{tabella1}) and Sect.~\ref{apun3}.} 
featuring operators of the type~\cite{jk,fo32},
%
\begin{equation}
{\bm \Psi}(x)=\sum_s\ \psi_s(x)\ \bm{a}_s
\end{equation}
that  Fermi indicates in a shortened form as in Eq.~\ref{capperi},
see Table~\ref{tabella3} for a description of terms and notations.
This relies on a complete basis of states $s$,  described by solutions 
$\psi_s(x)$ of the wavefunction, such that $\psi_s(x)\propto e^{-i E_s t/\hbar}$ with positive and negative energies $E_s$, normalized according to   Born $\int  |\psi_s(\vec{x})^2|\, d^3x=1$.
In this formalism,
$\bm{a}_s $ are adimensional 
operators that describe the elementary transitions  $\langle 0| \bm{a}_s  | s\rangle =1$ where we use
Dirac's notations to indicate a state of single fermion  $| s\rangle $ 
and the empty state $| 0\rangle $.  The conjugate operators, $\bm{a}_s^*$ and  ${\bm \Psi}^*$, 
describe the inverse transitions.

In the theory of electromagnetic interactions, the fields are only used in the 
$\bar{\bm \Psi} \gamma_\mu  {\bm \Psi}$ combination; in the theory of Fermi, for the first time, they are used alone; 
this remark corresponds to the comparison of Figures~\ref{fig1} and \ref{fig3}.

To avoid `ending up' in negative energy states, Fermi (following Dirac \cite{de29})  assumes
that all negative energy states of electrons and neutrinos are occupied and the Pauli principle applies. 
In formal terms, by separating the positive and negative energy states 
\begin{equation}
s=\left\{
\begin{array}{lll}
s_+ & \mbox{if} & E_s>0 \\
s_- & \mbox{if} & E_s<0 \\
\end{array}
\right.
\end{equation}
The assumed vacuum, known as the `Dirac sea'
 (Sect.~\ref{dirac})  is defined formally as
\begin{equation}
| \mbox{sea}_s\rangle =\prod_{s_-} {\bm{a} }_{s_-}^\dagger |0\rangle
\end{equation}
the product is over {\em all} particle states with negative energy.
Two independent Dirac seas are assumed by Fermi: one for electrons and one for neutrinos
(Sect.~\ref{cacaficio}); in order to shorten the notations, we will consider in the following
\begin{equation}
| \mbox{sea}\rangle = | \mbox{sea}_s\rangle \otimes | \mbox{sea}_\sigma\rangle 
\end{equation}
where, following the original notations 
we have denoted the electron states with $s$ and the 
neutrino states by $\sigma$ - see again Table~\ref{tabella3}.

%
%

\subsubsection{Implication for neutron decay}
To illustrate Fermi's scheme in the simplest way, 
let us consider the decay of the neutron\footnote{At the time, Fermi's theory was used to describe 
the change of state of neutrons {\em in nuclei};  the change of free neutrons was not yet observed.}
\begin{equation}
n\to p+ e + \nu
\end{equation}
The nucleon, according to Heisenberg \cite{isom}, is simply supposed to change its charge (isospin) state: from neutron to proton.
Fermi points out that, in this formalism, the creation of the electron can be described by the 
 following non-zero matrix element 
 \begin{equation}\label{erbv}
 \langle\mbox{sea} + s_{+} | \bm{\Psi}^*(x)  
| \mbox{sea}\rangle  =\varphi\cdot 
 \psi_{s_{+}}^*(x)
\end{equation}
and for the neutrino the consideration is identical,
\begin{equation}
 \langle\mbox{sea} + \sigma_{+} | \bm{\Phi}^*(x)  
| \mbox{sea}\rangle  = \varphi'\cdot 
\psi_{\sigma_{+}}^*(x)
\end{equation}
where, following the original notations $\bm{\Phi}$ 
 indicates second quantised field of the neutrinos 
 ($\varphi,\varphi'=\pm 1$ 
are  signs related to the order in which the states appear in the Dirac sea, that  
plays no role for the decay rate.)

Stated in plain words, it is formally possible to consider the creation of electrons and neutrinos.

\subsection{Details of the Fermi's Hamiltonian}\label{apun2}

\subsubsection{The current of relativistic particles}
The Hamiltonian function can be written as:
\begin{equation}
H_{int}=g\, \left[ \bm{Q}\ \bm{J}(x) + \bm{Q}^\dagger \bm{J}^\dagger(x) \right]
\end{equation}
and it includes:
1)~$g$, the Fermi constant;
2)~$Q=(\tau_1-i \tau_2)/2$, the isospin step operator, which allows to convert one atomic nucleus into another with a lower charge,
by turning a proton into a neutron\footnote{In our convention a proton (a neutron) 
is the eigenvector of the isospin operator $\tau_3$ with eigenvalue $+1$ ($-1$), where $\tau_{1,2,3}$ are the usual Pauli matrices. Thus, 
 $\tau_-$ decreases the charge of the nucleon and $\tau_+$ increases it, while  $(1+\tau_3)/2$ is the electric charge operator. 
Fermi consistently uses the opposite convention for isospin eigenvalues.}.
3)~the current $\bm{J}$ which describes the creation of an electron and a neutrino.
Fermi explains that {\sf\small \guillemotleft $x$ represents the coordinate of the heavy particle\guillemotright\footnote{This statement can be 
expanded with few manipulations. The  
hamiltonian of electromagnetic interactions with the scalar potential $\varphi$ is   
$H_{int}=q\int  \varphi(y)\  \bm{J}(y) \ dy$. For short-range nuclear interactions we  
replace $q\, \varphi(y) \to q\, \tilde{\varphi}(y)= g\, \delta(x-y)$; 
$x$ is the coordinate of the nucleus and $g$ a constant with dimensions volume$\times$energy.
We conclude that $\tilde{H}_{int}=q\int  \tilde{\varphi}(y) \  
 \bm{J}(y)\, dy=
g\times  \bm{J}(x) $.}.}


The discussion of the chosen form of the current $\bm{J}$ is interesting.
As a first approach, Fermi describes a current between scalar operators, simply given by
\begin{equation}
\bm{J} =\bm{\Psi} \bm{\Phi}
\end{equation}
where, to conform to the original notations, we indicate with
$\bm{\Psi}$ the field of electrons and $\bm{\Phi}$ that of neutrinos.
To describe {\em relativistic fermions} including the spin, 
Fermi gets inspiration from 
the form of the scalar coupling term $e\, V$ of electromagnetism, and thus writes
\begin{equation}
\bm{J}^0 = - \bm{\Psi}_1 \bm{\Phi}_2 + \bm{\Psi}_2 \bm{\Phi}_1
+\bm{\Psi}_3 \bm{\Phi}_4 - \bm{\Psi}_4 \bm{\Phi}_3
\end{equation}
He states that this term is the {\sf\small \guillemotleft scalar\guillemotright}  part  (i.e., the time-like component) 
of a combination of second quantisation fields which transform as a quadri-current.

\subsubsection{Translation into modern notation}
In modern terms, we can write the Fermi quadri-current as follows,
\begin{equation}
\bm{J}^\mu= \bm{\Psi}^t C^\dagger \gamma_5 \gamma^\mu \bm{\Phi}
\end{equation}
where $\gamma^5=i\gamma^0\gamma^1\gamma^2\gamma^3$ is the matrix now called chirality,
used by Pauli in 1936~\cite{pauli36}
and $C$ the charge conjugation
introduced by Kramers in 1937~\cite{kramers}, to which we will return in Appendix~\ref{apun3}.
This might surprise a modern reader, as in 1933 there was no mention of chirality or charge conjugation. On the other hand what really matters for Fermi's argument is simply Lorentz covariance of the Hamiltonian function, and the previous position, just as claimed by Fermi, guarantees it\footnote{Note that this current is compatible with parity conservation
and of the charge conjugation, defining
$\bm{\Psi}(x)\stackrel{\mbox{\tiny P}}{\to} \gamma_0\bm{\Psi}(x_{\mbox{\tiny P}}) $
and
$\bm{\Psi}(x)\stackrel{\mbox{\tiny C}}{\to} C \gamma_0^t\bm{\Psi}^*(x) $, and the same for
$\bm{\Phi}(x)$; to have a violation you need a current with two contrasting contributions \cite{ly}
as in the case of the $V-A$ current.}.
For the avoidance of doubt, we explicitly check
  the coherence between these two expressions, using the Dirac representation 
  of the $\gamma$ matrix. From 
\begin{equation}
\gamma^0=
\left(
\begin{array}{cc}
\sigma_0 &  0 \\
0 & - \sigma_0
\end{array}
\right), \qquad
\gamma^5=
\left(
\begin{array}{cc}
0 & \sigma_0  \\
\sigma_0 &  0
\end{array}
\right),\qquad
C=
\left(
\begin{array}{cc}
 0  & \varepsilon \\
\varepsilon & 0 
\end{array}
\right),
\end{equation}
where
$\sigma_0=\mbox{diag}(1,1)$, 
 $C=i\gamma^0\gamma^2$ 
and 
$\varepsilon=i\tau_2$
we get easily
\begin{equation}
\bm{J}^0 =
( \bm{\Psi}_1, \bm{\Psi}_2, \bm{\Psi}_3, \bm{\Psi}_4)
\left(
\begin{array}{cccc}
  0 & -1 & 0 & 0 \\
  1 & 0 & 0 & 0 \\
  0 & 0 & 0 & 1 \\
  0 & 0 & -1 & 0
  \end{array}
\right)
\left(
\begin{array}{c}
\bm{\Phi}_1 \\
\bm{\Phi}_2 \\
\bm{\Phi}_3 \\
\bm{\Phi}_4
   \end{array}
\right)
\end{equation}
which coincides with the Fermi expression.
Various authors proposed alternatives to Fermi's current, e.g.,
\cite{ku}, \cite{fierz} and \cite{racah}. 
The latter proposal, appeared after Majorana theory and due to Racah,  in modern notations reads 
$ (\bm{J}^0)^\dagger_R = \bar{\bm{\Psi}} \gamma^0 C ( \bar{\bm{\Phi}} )^t $.

\subsubsection{Observation}
Finally, let us consider the matrix elements of the lepton current.
From the conjugate current of the Fermi's theory,
$ (\bm{J}^0)^\dagger_F = \bar{\bm{\Psi}} \gamma^0 \gamma^5 C ( \bar{\bm{\Phi}} )^t
$, we find the matrix element of the current for 
the emission of a neutrino and an electron
\begin{equation}
{\mathcal J}_F=\bar{u}_e \gamma^0 \gamma^5 v_\nu
\end{equation}
where, in order to make the calculations more similar to the standard ones:\\
1)~we use the approximation of free particles,\\
2)~we adopt the covariant normalization of the spinors, $\bar{u} u=2m$,\\
3)~by convention we set $v= C (\bar{u})^t$ for the spinors.\\
The square modulus of the matrix element (that is, the 0-0 component of the `leptonic tensor') is
\begin{equation}
| \mathcal{J}_F |^2=\mbox{tr}[ ( \slashed{{p}} + m_e) \gamma^0 ( \slashed{k} + m_\nu ) \gamma^0 ]
= 4 (E \omega + \vec{p} \, \vec{k} + m_e m_\nu)
\end{equation}
where $p=(E,\vec{p}) $ and $k=({\omega},\ \vec{k})$ are the four-moments of the electron and the neutrino,
and we adopted for simplicity the convention $c=1$.
Let us see what changes using vector lepton current instead, 
$ (\bm{J}^0)^\dagger_V = \bar{\bm{\Psi}} \gamma^0 \bm{\Phi} $, i.e. the one that is typically used to talk about Fermi theory in class or in reviews.  
We have ${\mathcal J}_V=\bar{u}_e \gamma^0 v_\nu$, 
then $|{\mathcal J}_V^2|$ differs only for the 
term $m_e m_\nu$, which changes sign.
When the  neutrino mass is neglected, the two expressions  coincide. Compare with~\cite{racah}.
%

\subsection{Progress after Fermi's theory}\label{apun3}
We examine a few aspects of on an application developed to Wick of the Fermi's theory, and comment on 
two formal advances due to Majorana and Kramers.
\subsubsection{Wick's considerations}
Wick observes that Fermi's theory allow to describe 
a transition that at first sight seems entirely different, namely the electronic capture process
\begin{equation}
  p+ e \to n + \bar{\nu}
\end{equation}
where, in order for the reaction not to be prohibited by the conservation of energy,
one imagines that the proton is included in a suitable nucleus. Let us 
consider two matrix elements: that of the electron field, very similar to the previous case,
\begin{equation}
  \langle\mbox{sea} | \bm{\Psi}(x)
| \mbox{sea} + s_+ \rangle = \varphi''\cdot 
\psi_{s_+}(x)
\end{equation}
and that of the antineutrino, which is described by a `hole' in the Dirac sea
\begin{equation}
  \langle\mbox{sea} -\sigma_- | \bm{\Phi}(x)
| \mbox{sea} \rangle = \varphi'''\cdot \psi_{\sigma_-}(x)
\end{equation}
(the signs $\varphi'',\varphi'''=\pm 1$ are as in Eq.~\ref{erbv}).
Similar considerations apply to conjugate operators; in short, Fermi's theory 
leads to consider the so-called crossing symmetry, i.e., the relationship between different processes, described
from a prefixed Hamiltonian. This is an important feature, but 
evidently not exclusive  of quantum field theory: in fact, it was noted already 
in the context of `second quantisation'  (Sect.~\ref{cdb}).

%

\subsubsection{Majorana's symmetric quantisation}
Pauli and Weisskopf \cite{pw}
proved that the scalars can be quantized consistently, which  is enough to refute the claim of Dirac that only fermions can be quantized, but 
leaves open the question of what to do with fermions.
The first person to propose the modern quantisation procedure
for fermionic fields (also told canonical quantisation) has been 
Ettore Majorana; we resume his proposal emphasising the differences with the previous position,
rather than following in detail his argumentation. 
Majorana  notes that it is possible to choose the four Dirac matrices $ 
  \gamma_\mu $ so that they are purely imaginary; thus, 
   Lorentz transformations 
  $\psi'=\Lambda(\omega)\psi$  are represented by 
 the real matrices
$\Lambda(\omega)=\exp(-i \omega_{\mu\nu} \Sigma^{\mu\nu}/4)$
as the spin matrices 
  $\Sigma^{\mu\nu}=i [\gamma^\mu,\gamma ^\nu]/2$
are imaginary.

Therefore, is possible to consider a real (or as we say today, Majorana's) fermionic field
%
\begin{equation}
{\bm \Psi}^{\mbox{\tiny real}}_a(x)=\sum_{s=s_+}\left( 
\psi_s(x)\ \bm{a}_s +  \psi_s^{\mbox{\tiny *}}(x)\ \bm{a}_s^\dagger
\right)
\end{equation}
where we highlight that the sum is only over the positive energies. 
This resembles closely the quantised field of the photon: in other words, at this point the analogy between radiation and matter is also formally complete.
With two real fields, we can form a complex one, that can represent a charged particle,
\begin{equation}
{\bm \Psi}^{\mbox{\tiny cplx}}_c(x)=\frac{ {\bm \Psi}^{\mbox{\tiny real}}_a(x)+ i\ {\bm \Psi}^{\mbox{\tiny real}}_b(x)}{\sqrt{2}}
\end{equation}
that we can write explicitly as
\begin{equation}
{\bm \Psi}^{\mbox{\tiny cplx}}_c(x)=\sum_{s=s_+}
\left(   \psi_s(x)\ \bm{c}_s +  \psi_s^{\mbox{\tiny *}}(x)\ \bm{\bar{c}}_s^\dagger
 \right)
\end{equation}
where
\begin{equation}
\bm{c}_s=\frac{\bm{a}_s+ i \bm{b}_s} {\sqrt{2}} \neq \bm{\bar{c}}_s=\frac{\bm{a}_s-  i \bm{b}_s} {\sqrt{2}} 
\end{equation}
There is no need to talk of `negative energies': 
for real fields and for complex fields,  only
wave functions with positive frequency are used,
%
\begin{equation}
\psi_s(x)\propto e^{ -i 2\pi \nu_s t}\mbox{ where }E_s=h \nu_s>0\  \forall s
\end{equation}
In this way, the quantisation procedure is symmetric between particles and antiparticles
and vacuum is simply the state in which there is no fermion -
no need to introduce the Dirac sea,  just as in 
modern QFT treatment for fermions\footnote{To be sure, the  expression of the complex field  in a generic  
  representation of Dirac matrices is
%
\begin{equation}
{\bm \Psi}_{\mbox{\tiny QFT}}(x)=\sum_{s=s_+}
\left(   \psi_{s}^{(a)}(x)\ \bm{a}_s +  \left[ \psi_s^{(b)} (x) \right]^{\mbox{\tiny C}}\, \bm{b}_s^\dagger
\right)
\end{equation}
The sum is only over positive energy states.
We  changed the names of the creation and destruction operators to  
conform to usual notations: $a$ indicates particles,
$b$  antiparticles. We define $\psi^{\mbox{\tiny C}}= C(\bar{\psi})^t$, where 
$C$ is the charge conjugation matrix \cite{kramers} introduced shortly 
after the publication of the Majorana's work~\cite{maj}.
The matrix $C$,  unitary and with determinant 1,
obeys the condition $C\ \gamma_\mu^t= - \gamma_\mu C$; its 
existence is ensured by Pauli's theorem \cite{pauli36} on unicity of Dirac matrices.
The vacuum takes the simplest possible form: it is just the state $|0\rangle$ where there is no particle.
The single particle states are those of the Fock construction, based on the above ingredients
\begin{equation}
\left\{
\begin{array}{lc}
|a_s\rangle= \bm{a}_{s}^\dagger |0\rangle & \mbox{if particle} \\[1ex]
|b_s\rangle= \bm{b}_{s}^\dagger |0\rangle & \mbox{if antiparticle}\\
\end{array}
\right.
\end{equation}
Thus, the field
${\bm \Psi}_{\mbox{\tiny QFT}}$  describes the disappearance of a particle from the initial state,
or the creation of an antiparticle in the final state
\begin{equation}
\langle 0 | {\bm \Psi}_{\mbox{\tiny QFT}}(x) |a_s\rangle= \psi_{s}^{(a)}(x) \quad ; \quad
\langle b_s| {\bm \Psi}_{\mbox{\tiny QFT}}(x) |0\rangle= \left[ \psi_s^{(b)} (x) \right]^{\mbox{\tiny C}}
\end{equation}
and similarly  for conjugated operator $\bm{\Psi}^\dagger_{\mbox{\tiny QFT}}$.
Evidently, the modern quantised  fermion fields coincides, up to 
 a formal generalisation of minor significance - the introduction of $C$ matrix - with Majorana symmetric quantisation.}.


\medskip

Let us conclude with an important remark on neutrinos:
While complex fields are needed to describe the 
charged fermions (e.g., electrons and positrons),  for neutral particles - and in particular for neutrinos -  real fields are sufficient.
This hypothesis, due  again to Majorana,  expresses a new concept of neutrinos which {\em differs} from the one of Fermi's theory:
for a very early discussion,   which preceded the important discovery of the $V-A$ structure of the interactions,
see~\cite{racah}.
Majorana hypothesis is part of the simplest conceivable extension of the standard model of interactions and 
particles that is consistent with all accepted observations: in fact, it explains the evidence of neutrino masses without postulating 
new light particles, and moreover, it leads to important, potentially testable predictions.
Actually, Majorana neutrinos are 
at the forefront of current particle physics; see~\cite{papa} for a historical outline, \cite{mois} for a 
recent theoretical discussion and \cite{tois} for an extensive review.

\newpage

{\printendnotes }

\newpage

{


}
 
\newpage
 
\parskip0.5ex
\footnotesize
\tableofcontents

\end{document}